\def\b{\begin{equation}}
\def\e{\end{equation}}
\def\balll{\begin{array}{lll}}
\def\ea{\end{array}}
\def\bea{\begin{eqnarray}}
\def\eea{\end{eqnarray}}

\documentclass[floatfix,twocolumn,showpacs,preprintnumbers,amsmath,amssymb,eqsecnum,aps]{revtex4}
\usepackage{graphicx}

\begin{document}
\title{Scalar radiation emitted from a rotating source
around a Reissner-Nordstr\"om black hole}

\author{Lu\'\i s C. B. Crispino}
\email{crispino@ufpa.br}
\affiliation{Faculdade de F\'\i sica, Universidade Federal do
Par\'a, 66075-110, Bel\'em, PA,  Brazil}

\author{Andr\'e R. R. da Silva}
\email{dasilva@ift.unesp.br}
\affiliation{Instituto de F\'\i sica Te\'orica, Universidade Estadual Paulista,
Rua Pamplona 145, 01405-900, S\~ao Paulo, SP, Brazil}

\author{George E. A. Matsas}
\email{matsas@ift.unesp.br}
\affiliation{Instituto de F\'\i sica Te\'orica, Universidade Estadual Paulista,
Rua Pamplona 145, 01405-900, S\~ao Paulo, SP, Brazil}

\date{\today}
\begin{abstract}
We investigate the radiation emitted from a scalar source in circular
orbit around a Reissner-Nordstr\"om black hole. Particle and energy
emission rates are analytically calculated in
the {\em low-} and {\em high-}frequency regimes and shown to be in full
agreement with a numerical calculation. Our investigation is connected
with the recent discussion on the validity of the {\em cosmic
censorship conjecture} in the quantum realm.
\end{abstract}

\maketitle

\section{Introduction}
\label{sec:Introduction}

This is currently under investigation if asymptotically
flat spacetimes endowed with some {\em suitable} matter
content evolved through Einstein equations from some
generic initial conditions can give rise to spacetimes
with {\em naked singularities} (see, e.g.,
Refs.~\cite{shell,ES79,C84,JD93,C93,HHM04} and references therein).
According to the cosmic censorship conjecture (CCC)
put forward by Penrose in 1969 such naked singularities should
never occur~\cite{P69}. The validity of the various versions of the CCC
is presently source of intense debate (see, e.g.,
Refs.~\cite{grw,clarke,grqc,penrose} for comprehensive accounts).

In contrast to naked singularities, the ones present in the
interior of black holes are dressed by event horizons. According to the
uniqueness theorems, all stationary black hole
solutions of the Einstein-Maxwell equations are uniquely determined
by the gravitational mass $M$, electric charge $Q$, and angular
momentum $J$ satisfying the relationship $ M^2 \geq Q^2 + (J/M)^2 $.
(We assume natural units $c=G=\hbar=1$ unless stated otherwise.)
If we were able to violate the inequality above by {\em overcharging}
and/or {\em overspinning} a black hole, then we would
end up with a naked singularity characterized by $ Q^2 + (J/M)^2 > M^2$.
Because stationary black holes are stable under linear
perturbations~\cite{vish,price,kay,whit}, they would be a good testing
ground for such a search. In 1974, Wald wondered whether or not an
extreme black hole $M^2 = Q^2+(J/M)^2$ could be overcharged or
overspined by the absorption of a {\it classical} particle with some
proper angular momentum and/or charge~\cite{gedan}. He eventually
shows that in order to overcome the gravitational barrier,  the classical
particle must have just enough energy to compensate any angular momentum
and/or charge carried into the hole (see also Ref.~\cite{tristan}). As a
result, the black hole constraint $ M^2 \geq Q^2 + (J/M)^2 $ would be
preserved  and the CCC upheld. This illustrates the current belief
that {\em classical} mechanisms are not expected to undress the singularity
hidden in the interior of black holes~\cite{lsss}.

On the other hand, it is well known that  quantum mechanics can
sometimes jeopardize classical assumptions leading to opposite
conclusions with respect to the ones obtained in the classical
realm. Working in the context of {\em quantum field theory in
curved spacetime} (QFTCS) Hawking, e.g., was able to show that black
holes can evaporate~\cite{hawk_rad} in contrast to the overspread
general-relativistic belief that this would be impossible.
It seems natural, then, to inquire as to whether quantum mechanics
would have something to add to the CCC issue.
An investigation on these lines was recently performed by
Ford and Roman~\cite{ford1} who analyzed the possibility of
producing a naked singularity by injecting negative energy into an extreme
charged black hole. They concluded that the positive energy flux which
always follows the negative one would render the CCC true.
More recently~\cite{MS07} it was wondered whether or not a scalar
wave with {\em small energy} but {\em large enough angular momentum}
could tunnel through the gravitational scattering potential of a nearly
extreme macroscopic Reissner-Nordstr\"om black hole, $|Q|/M \lesssim 1$,
whereby it would acquire enough angular momentum to
overspin, $Q^2+(J/M)^2> M^2 $, and therefore challenge the
CCC. The generalized second law of thermodynamics could still be
preserved if the initial entropy of the black hole were
carried away by the degrees of freedom of some final {\em debris}
assuming that naked singularities are unstable.
Although a first approximation calculation shows
that this is possible, a further analysis performed by
Hod~\cite{H08} suggests that the backreaction on the background
spacetime due to the wave angular momentum  would preclude it of being
absorbed rendering the CCC true again. This is interesting
because up to the present knowledge there is no mandatory
reason to preclude the formation of naked singularities in the quantum
realm. It is largely believed that {\em quantum gravity} should be able
to unveil the physical structure of these ``entities" making them
nonsingular  and recovering the spacetime predictability.  Indeed, quite
recently Richartz and Saa~\cite{RS08} have argued that the quantum
overspinning mechanism can be rendered true by replacing the scalar
by a fermionic field. This is so because in contrast to the scalar field
the fermionic one would not be superradiated away. Probably a final
veredictum on the validity of CCC will not be possible before a full
quantum formulation of gravity is achieved. In the meanwhile, however,
QFTCS can bring us useful insights about this issue.
\begin{figure}[b]
\includegraphics[width=8cm]{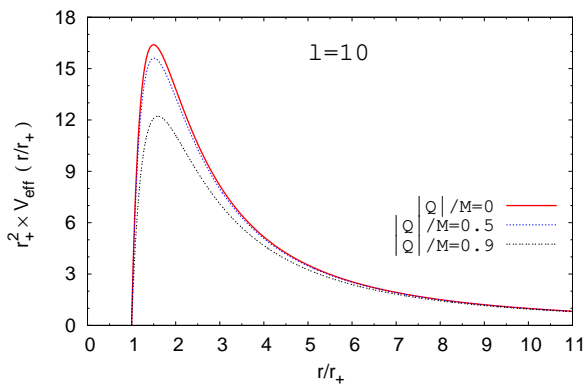}
\caption{The effective scattering potential for $|Q|/M = 0, 0.5~{\rm and}~
0.9$ is plotted for $l=10$ as a function of $r/r_+$. The larger the $|Q|/M$
the smaller the $V_{\rm eff} (r/r_+)$.}
\label{figurepotential}
\end{figure}

Here we realize a production mechanism of {\em ingoing} and {\em
outgoing} scalar particles {\em towards} and {\em from} the
Reissner-Nordstr\"om black hole, respectively, by considering a
monopole in circular orbit around the hole coupled to a massless
Klein-Gordon field. Refs.~\cite{MS07,H08} assume scalar particles
beamed towards the hole but do not mention how they would be
produced. Because here we will be mostly interested in waves with
small energy and angular momentum in comparison to the hole mass, no
spacetime backreaction considerations  are in order. This
investigation is not only interesting in connection with
Refs.~\cite{MS07,H08} but also fulfills a gap in the literature on
the so called {\em gravitational synchrotron radiation} initiated in
the 70's by Misner and collaborators~\cite{Misneretal} and followed
up to these days by a number of authors~\cite{Lemosetal}, since
scalar radiation emitted from sources around black holes when they are
endowed  with electric charge has not been considered yet. The usual flat 
spacetime synchrotron signature can be seen when the monopole moves 
fast enough.
We work in the context of standard QFTCS (see Refs.~\cite{birrel,fulling}
for comprehensive accounts). Because of the difficulty to express
the solution of some differential equations which we deal with in
terms of known special functions, our computations are performed (i)
numerically but without further approximations and (ii) analytically
but restricted to the low- and high-frequency regimes. The paper is
organized as follows. In section \ref{sec:emissionrates} we present
the general formulas for the emission rate and radiated power of
scalar particles from the monopole source in circular orbit around
the Reissner-Nordstr\"om black hole. In section \ref{sec:lowandhigh}
we present analytic results in the low- and high-energy regimes. In
section~\ref{sec:results}, the analytic results obtained in the
previous section are plotted against a full numerical calculation and
shown to agree. We also compute the amount of the emitted radiation
which reaches asymptotic observers rather than being absorbed by the
hole. Section~\ref{sec:finalremarks} is dedicated to our
final remarks.

\section{Emission rates and radiated powers}
\label{sec:emissionrates}

The line element of a Reissner-Nordstr\"om black hole with mass $M$
and electric charge $|Q|\leq M$ can be written as~\cite{grw}
\begin{equation}
     ds^2 =
        f(r) dt^2 - f(r)^{-1} dr^2 - r^2 \left( d\theta^2 + \sin^2
\theta d\varphi^2 \right),
  \label{34}
\end{equation}
where
\begin{equation}
     f(r) \equiv(1-r_+/r)(1-r_-/r)
  \label{34.5}
\end{equation}
and
$
r_\pm \equiv M \pm \sqrt{M^2-Q^2}.
$
Outside the outer event horizon, i.e. for $r > r_+$, we have a global
timelike isometry generated by the Killing field $\partial_t$.

Now we introduce a free massless scalar field $\Phi=\Phi (x^\mu)$
satisfying $\Box \Phi = 0$. The corresponding field operator can be
expanded in terms of creation ${a^{\alpha \dagger}_{\omega l m}}$
and annihilation $a^\alpha_{\omega l m}$ operators as
\begin{equation}
        \hat \Phi(x^\mu) =
        \sum_{\alpha = \leftarrow}^{\rightarrow}
        \sum_{l=0}^{\infty}
        \sum_{m=-l}^{l}
        \int_0^{\infty} d\omega
        [u^\alpha_{\omega l m}(x^\mu) a^\alpha_{\omega l m} + {\rm H.c.}],
   \label{10.2}
\end{equation}
where the normal modes are written as
\begin{equation}
u^\alpha_{\omega l m} =
\sqrt{\frac{\omega}{\pi}} \frac{\psi^\alpha_{\omega l}(r)}{r}
Y_{lm}(\theta,\varphi) e^{-i\omega t}
\label{18}
\end{equation}
and are assumed to be orthonormalized according to
the Klein-Gordon inner product~\cite{birrel}.
Here $\omega \ge 0$ and $l \ge 0$, $m \in [-l,l]$ are frequency
and angular momentum quantum numbers, respectively, and
$\alpha= \leftarrow (\rightarrow )$
labels ingoing (outgoing) modes. $Y_{lm}(\theta,\varphi)$ are the usual
spherical harmonics. $\psi_{\omega l}^{\leftarrow} (r)$
and $\psi_{\omega l}^{\rightarrow}(r)$ are associated with purely
incoming modes from the past null infinity ${\cal J}^-$ and outgoing
from the past white-hole horizon ${\cal H}^-$, respectively.
$\psi^\alpha_{\omega l}$ satisfies
\begin{equation}
     \left[
     - f(r) \frac{d}{dr} \left( f(r) \frac{d}{dr} \right) + V_{\rm eff}(r)
     \right] \psi^\alpha_{\omega l}(r) =  \omega^2 \psi^\alpha_{\omega l}(r),
   \label{40}
\end{equation}
where
\begin{equation}
     V_{\rm eff}(r) =
         \left(
        1- \frac{2M}{r} +\frac{Q^2}{r^2}
        \right)
        \left(
        \frac{2M}{r^3} -\frac{2 Q^2}{r^4}  + \frac{l(l+1)}{r^2}
        \right)
   \label{41}
\end{equation}
is the effective scattering potential (see, e.g., Ref.~\cite{cm} for
more detail). A plot of the scattering potential can be found in
Fig.~\ref{figurepotential}. The larger the $l$
the larger the $V_{\rm eff}$ because of the centrifugal barrier. By
performing the coordinate transformation
\begin{equation}
      x \equiv
        y + \frac{ (y_+)^2 \ln\left| y-y_{+} \right| -
        (y_-)^2\ln\left| y-y_{-}
        \right| }{y_+ - y_-},
   \label{52}
\end{equation}
where
$y \equiv r/2M$ and $y_{\pm} \equiv r_{\pm}/2M$,
Eq.~(\ref{40}) can be cast in the form
\begin{equation}
     (- {d^2}/{dx^2} + 4M^2 V_{\rm eff}[r(x)]) \psi^\alpha_{\omega l}(x)
=
        4M^2 \omega^2 \psi^\alpha_{\omega l}(x).
\label{53}
\end{equation}
\begin{figure}[t]
\includegraphics[width=8cm]{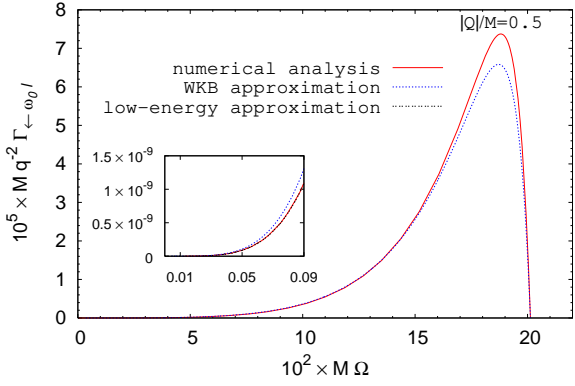}
\caption{The numerical result for $\Gamma_{\leftarrow \omega_0 l}$
with $l=m=5$ is shown assuming a black hole with $|Q|/M = 0.5$.
The internal box is a zoom of the $\Omega M \ll 1$ region and shows
the excellent agreement obtained with our low-energy formulas. The
numerical and low-energy results are superimposed and cannot be
distinguished with the present resolution. We also plot in this region
the result obtained with the WKB method to make it clear that it captures
the  qualitative behavior in the low-energy region as well. Finally,
we emphasize the very nice quantitative approximation
provided by the WKB method in the $R_S \approx r_{\rm ph}$ region
($\Omega M \approx 0.2$).
}
\label{crin}
\end{figure}
Accordingly, the creation and annihilation operators satisfy the
simple commutation relations
\begin{equation}
[ a^\alpha_{\omega l m}, {a^{\alpha' \dagger}_{\omega' l' m'}} ] =
    \delta_{\alpha \alpha'} \delta_{l l'}
    \delta_{m m'}\delta(\omega - \omega'),
   \label{10.3}
\end{equation}
where the state $|0\rangle$, defined by $a^\alpha_{\omega lm}|0\rangle = 0$
for every $\alpha, \omega, l$ and $m$, is denominated {\em Boulware vacuum}.
Close  ($x<0, |x| \gg 1 $) to and far away ($x \gg 1 $) from the horizon we
have
\begin{equation}
     \psi_{\omega l}^{\leftarrow}(x) \approx \frac{1}{2 \omega}
       \left\{
          \begin{array}{lc}
            {\cal T}_{\omega l}^{\leftarrow} e^{-2iM\omega x}
            \;\;\; (x< 0, |x| \gg 1) \\
            \\
            2(-i)^{l+1} M\omega x {h_l^{(1)}(2M\omega x)}^*
            \\
            + 2 i^{l+1} {\cal R}_{\omega l}^{\leftarrow} M\omega x h_l^{(1)}(2M\omega x)
            \;\;\; (x \gg 1)
          \end{array} \right.
   \label{68}
\end{equation}
and
\begin{equation}
     \psi_{\omega l}^{\rightarrow}(x) \approx \frac{1}{2 \omega}
        \left\{
          \begin{array}{l}
         e^{2iM\omega x} + {\cal R}_{\omega l}^{\rightarrow} e^{-2iM\omega x}
         \; (x < 0\;, |x| \gg 1) \\
         \\
         2 i^{l+1}
         {\cal T}_{\omega l}^{\rightarrow} M\omega x
         h_l^{(1)}(2M\omega x)
         \; (x \gg 1)
          \end{array}
        \right.
   \label{51}
\end{equation}
\begin{figure}[t]
\includegraphics[width=8cm]{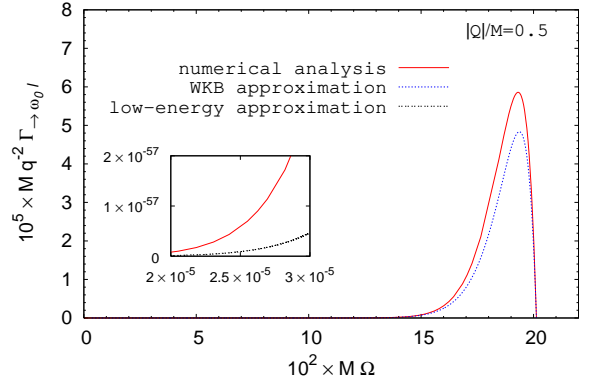}
\caption{The numerical result for $\Gamma_{\rightarrow \omega_0 l}$
with $l=m=5$ and $|Q|/M = 0.5$ is plotted. We can see
the convergence of the numerical calculation with
(i) low-energy analytical  results for $\Omega M \ll 1$
and (ii) the WKB method for $R_S \approx r_{\rm ph}$ 
($\Omega M \approx 0.2$).}
\label{crout}
\end{figure}
Here $ \left| {\cal R}_{\omega l}^{\alpha} \right|^2 $ and $ \left|
{\cal T}_{\omega l}^{\alpha} \right|^2 $  are the reflection and 
transmission coefficients,
respectively, satisfying the usual probability conservation
equation: $ \left| {\cal R}_{\omega l}^\alpha \right|^2 + \left|
{\cal T}_{\omega l}^\alpha \right|^2 = 1 $ and $h_l^{(1)}(2M \omega
x)$  is the spherical Hankel function. Note that $h_l^{(1)}(x)
\approx (-i)^{l+1} \exp(ix)/x $ for $|x| \gg 1$.

Now let us consider a monopole
\begin{equation}
j (  x^{\nu} )  =\dfrac{q}{\sqrt{-g}\, u^{0}}\delta\left(
r-R_{S}\right)  \delta\left(  \theta-\pi/2\right)  \delta\left(
\varphi-\Omega t\right)
\label{jS}
\end{equation}
describing a scalar source in uniform circular motion
at the equatorial plane of the Reissner-Nordstr\"om black hole, i.e.,
$\theta=\pi/2$, with $r=R_{S}$ and angular velocity
$\Omega\equiv d\phi/dt= {\rm const} >0$  as defined by
asymptotic static observers. Here $ g \equiv {\rm det} (g_{\mu \nu}) $
and
\begin{equation}
u^{\mu}\left(  \Omega,R_{S}\right)  =
(f \left( R_{S}\right)  -R_{S}^{2}\Omega^{2} )^{-1/2}
( 1 ,0,0,\Omega)
\label{uS}
\end{equation}
is the four-velocity of the source. By assuming that the source
is free of interactions other than the gravitational one, we obtain
that
\begin{equation}
\Omega = \sqrt{M/R_S^3 - Q^2/R_S^4~},
\label{Kepler}
\end{equation}
where
\begin{equation}
R_S > r_{\rm ph} = (3M+\sqrt{9M^2-8Q^2})/2.
\label{photon}
\end{equation}
Here $r_{\rm ph}$ is the radius of the null circular geodesic
and defines the innermost limit to timelike geodesic circular orbits.
We note, moreover, that we have normalized the source $j(x^\mu )$
in Eq.~(\ref{jS}) by requiring that $\int d\sigma j(x^\mu ) = q = {\rm const}$,
where $d \sigma$ is the proper three-volume element orthogonal to $u^\mu$.

Next let us minimally couple the source to the field
through the interaction action
\begin{equation}
\hat{S}_{I}=
\int d^4 x \sqrt{-g}~j \hat\Phi.
\label{acaoint}
\end{equation}
From this we can interpret $q$ in Eq.~(\ref{jS}) as a coupling constant between
source and field.
Then the emission amplitude at the tree level of one
scalar particle with quantum numbers $(\alpha,\omega ,l,m)$
into the Boulware vacuum is given by
\begin{eqnarray}
\mathcal{A}^{\rm em}_{\alpha \omega lm}
& = & \langle \alpha \omega lm
      | i \hat{S}_I |
      0 \rangle
\nonumber
\\
& = & i \int d^{4}x~\sqrt{-g}~j(x^\mu)~ u^{\alpha *}_{\omega lm}.
\label{emissao}%
\end{eqnarray}
Note that for sources in constant circular motion the amplitude
$\mathcal{A}^{\rm em}_{\alpha \omega  l m }$ is proportional to
$\delta (\omega - \omega_0 )$ where we have defined
$\omega_0 \equiv m \Omega$. Hence the frequency  of the emitted
particles is constrained by the relation $\omega = \omega_0$. In
particular, since $\Omega >0$, no waves with $m \leq 0$ are emitted.
The emission rate $\Gamma_{\alpha \omega_0 l}$ and corresponding
emitted power $W_{\alpha \omega_0 l}$ of particles with
quantum numbers $(\alpha, \omega_0, l)$ ($l\geq 1$)
are given by
\begin{eqnarray}
 \Gamma_{\alpha \omega_0 l}
& = &
\int^{+ \infty}_0 d\omega \,
{\left| {\cal A}^{\rm em}_{\alpha \omega l m} \right| ^2}/{T}
\nonumber \\
& = &
2 q^2 \omega_0~(f(R_S) - R_S^2 \Omega^2)~
| \psi^\alpha_{\omega_0 l}(R_S)/R_S |^2
\nonumber \\
&\times&
|Y_{lm}(\pi/2,0) |^2
\label{Gammaem}
\end{eqnarray}
and
\begin{eqnarray}
W_{\alpha  \omega_0 l}
& = &
\int^{+ \infty}_0 d\omega \, \omega \;
{\left| {\cal A}^{\rm em}_{\alpha \omega l m} \right| ^2}/{T}
\nonumber \\
& = &
2 q^2 \omega_0^2~(f(R_S) - R_S^2 \Omega^2)~
| \psi^\alpha_{\omega_0 l}(R_S)/R_S |^2
\nonumber \\
&\times&
|Y_{lm}(\pi/2,0) |^2,
\label{Wem}
\end{eqnarray}
respectively, where $T=2\pi \delta(0)$ is the total time
as measured by asymptotic observers~\cite{IZ}.
Note also that $Y_{lm}(\pi/2,0) = 0$ if $l+m$ is odd and
\begin{equation}
        |Y_{l m}(\pi/2,0)|^2 =
\frac{2l+1}{4\pi}\frac{(l+m-1)!!(l-m-1)!!}{(l+m)!!(l-m)!!}
\label{Y}
\end{equation}
if $l+m$ is even~\cite{GR}. We have defined $n!! \equiv n(n-2)\cdots 1$ if
$n$ is odd,  $n!! \equiv n(n-2)\cdots 2$ if $n$ is even and $(-1)!! \equiv 1$.
Moreover, note that if we had
chosen the Unruh or Hartle-Hawking vacua rather than the Boulware
one then Eqs.~(\ref{Gammaem})-(\ref{Wem})
would be associated with the {\em net} emitted radiation
since the absorption and stimulated emission rates
(which are induced  by the presence of thermal fluxes) are
the same.

The total emission rate $\Gamma^{\rm total}$ and radiated
power $W^{\rm total}$ are obtained by summing on the quantum numbers
$\alpha,  l, m$ in Eqs.~(\ref{Gammaem}) and~(\ref{Wem}), accordingly.
The total particle and energy rates which escape to infinity
are
\begin{equation}
\Gamma^{\rm obs} = \sum_{l=1}^{\infty} \sum_{m=1}^{l}
                   (|{\cal T}_{\omega_0 l}^\rightarrow|^2
                   \Gamma_{\rightarrow \omega_0 l} +
                   |{\cal R}_{\omega_0 l}^\leftarrow|^2
                   \Gamma_{\leftarrow \omega_0 l})
\label{Gammaobs}
\end{equation}
and
\begin{equation}
W^{\rm obs} =      \sum_{l=1}^{\infty} \sum_{m=1}^{l}
                   (|{\cal T}_{\omega_0 l}^\rightarrow|^2
                   W_{\rightarrow \omega_0 l} +
                   |{\cal R}_{\omega_0 l}^\leftarrow|^2
                   W_{\leftarrow \omega_0 l}),
\label{Wobs}
\end{equation}
respectively. Here we note that
${\cal T}^\leftarrow_{\omega_0 l} = {\cal T}^\rightarrow_{\omega_0 l}$.
This guaranties that
$ |{\cal R}^\leftarrow_{\omega_0 l}| = |{\cal R}^\rightarrow_{\omega_0 l}| $.
Note, however, that  ${\cal R}^\leftarrow_{\omega_0 l}$ and
${\cal R}^\rightarrow_{\omega_0 l}$
will in general differ by a phase  (in contrast to
${\cal T}^\leftarrow_{\omega_0 l}$ and ${\cal T}^\rightarrow_{\omega_0 l}$).
\begin{figure}[t]
\includegraphics[width=8cm]{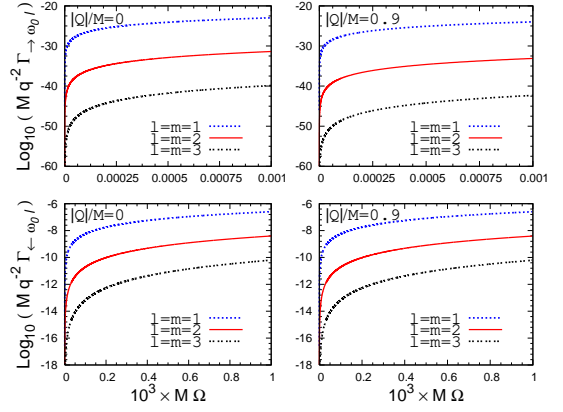}
\caption{The particle emission rates
$\Gamma_{\alpha \omega_0 l}$
($\alpha = \leftarrow, \rightarrow$) are plotted
in the low-frequency regime as a function of the
source angular velocity $\Omega$ for different values of $l$ ($m=l$).
$\Gamma_{\leftarrow \omega_0 l} $
is seen to be larger than $\Gamma_{\rightarrow \omega_0 l} $.
The larger the $l$ and $|Q|/M$ the smaller the
$\Gamma_{\alpha \omega_0 l}$.
}
\label{lowresponse}
\end{figure}
\begin{figure}[b]
\includegraphics[width=8.5cm]{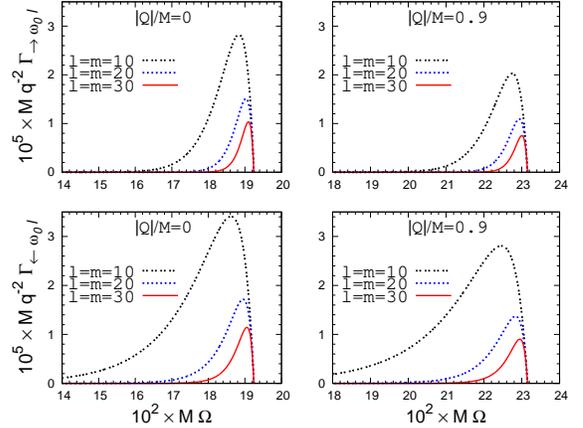}
\caption{The particle emission rates
$\Gamma_{\alpha \omega_0 l}$ are plotted
in the high-frequency regime for different values of $l$ ($m=l$).
The qualitative behavior of $\Gamma_{\alpha \omega_0 l}$
are much like the ones observed in the low-frequency regime
with one exception, namely, $\Gamma_{\rightarrow \omega_0 l} $
becomes larger than $\Gamma_{\leftarrow \omega_0 l} $
for $R_S \approx r_{\rm ph}$.
The existence of charge in the black hole makes
$\Gamma_{\alpha \omega_0 l}$ to decrease.
The figures are
plotted up to the last timelike  geodesic circular orbit.}
\label{highresponse}
\end{figure}
\section{Low- and high-energy solutions}
\label{sec:lowandhigh}

Now, in order to calculate the physical observables
given by Eqs.~(\ref{Gammaem})-(\ref{Wem}) and~(\ref{Gammaobs})-(\ref{Wobs})
we must work out the functions
$\psi^\alpha_{\omega l}(r)$. We exhibit approximate
low- and high-frequency solutions which are going to be used
in the next section in conjunction with a full numerical
calculation designed to cover the whole frequency range.

\subsection{Low-energy solutions}
\label{sec:low}

The low-frequency solution for $\psi^\alpha_{\omega l}(r)$
has been already worked out in Ref.~\cite{cm} and can be cast
(up to an arbitrary phase) in the form
\begin{equation}
      \psi^\rightarrow_{\omega l}(r)=
               \frac{ -4 i M  y_{+}~y~Q_{l}[z(y)]}{ y_+ - y_- }
   \label{uI}
\end{equation}
and
\begin{equation}
      \psi^\leftarrow_{\omega l}(r)=
       \frac{2^{2l+1} (-i)^{l+1} (l!)^3 M^{l+1} (y_+ - y_-)^l
       \omega^{l}~y P_{l}[z(y)]}{(2l+1)!~(2l)!},
   \label{uII}
\end{equation}
where
\begin{equation}
     z (y) \equiv
        \frac{2y - 1}{y_+ - y_-}.
   \label{44}
\end{equation}
One can also obtain
\begin{equation}
{\cal T}^\rightarrow_{\omega l} =
 \frac{ 2^{2l+2} (-i)^{l+1} y_+ (y_+ -y_-)^l (l!)^3 (M\omega)^{l+1}}
      { (2l+1)! (2l)!}
\label{tauout}
\end{equation}
in the low-frequency regime.  We recall that
${\cal T}^\leftarrow_{\omega l} = {\cal T}^\rightarrow_{\omega l}$.
Eq.~(\ref{tauout}) was used in Ref.~\cite{MS07} to calculate the
probability $|{\cal T}^\rightarrow_{\omega l}|^2$
of a wave to tunnel into the black hole assuming a fixed
Reissner-Nordstr\"om effective scattering potential. The larger the
black hole mass and charge in comparison with the wave energy and
angular momentum the better the static potential approximation.
Here we consider large enough black holes in order to neglect
spacetime backreaction effects.

\subsection{High-energy solutions}
\label{sec:high}
\begin{figure}[t]
\includegraphics[width=8cm]{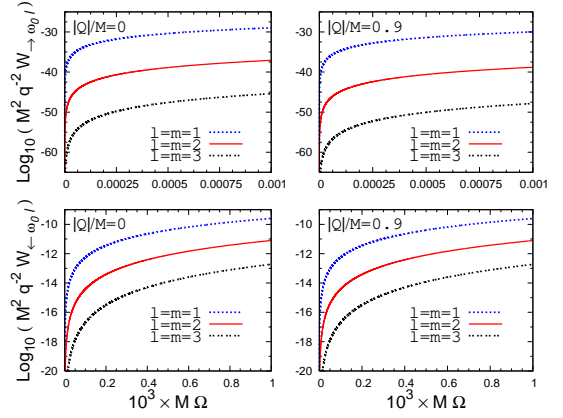}
\caption{The radiated powers
$W_{\alpha \omega_0 l}$
($\alpha = \leftarrow, \rightarrow$) are plotted
in the low-frequency regime as a function of the
source angular velocity $\Omega$ for different values of $l$ ($m=l$).
The qualitative behavior of $W_{\alpha \omega_0 l}$
and $\Gamma_{\alpha \omega_0 l}$ are similar to each other
in the low-frequency regime in contrast to what we see in the
high-frequency one.}
\label{lowpower}
\end{figure}

\begin{figure}[b]
\includegraphics[width=8.5cm]{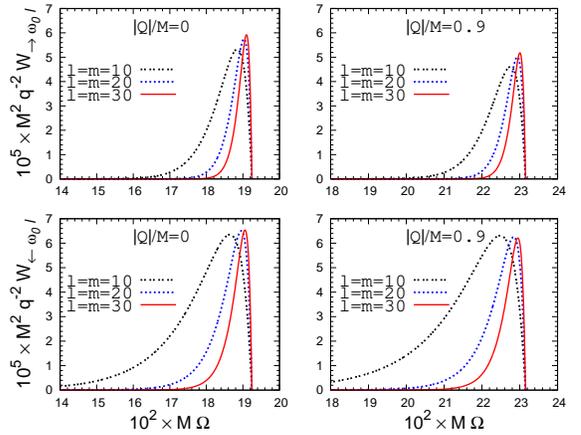}
\caption{The radiated powers
$W_{\alpha \omega_0 l}$
($\alpha = \leftarrow, \rightarrow$) are plotted
in the high-frequency regime. Note that depending on the
value of $\Omega$ different $l$ values contribute the
most. }
\label{highpower}
\end{figure}
A good approximation of $\psi^\alpha_{\omega l}(r)$ for high
energies can be obtained by using the WKB approximation (see, e.g.,
Ref.~\cite{wkb}). To do so, it is worth noting that Eq.~(\ref{53})
resembles the one-dimensional Schr\"odinger equation. By considering
the effective energy $\omega^2$ lower than the peak of
$V_{\rm{eff}}(x)$, one has two distinct situations. The first one is
characterized by $V_{\rm{eff}}(x) < \omega^2$ which is valid in the
intervals $(-\infty ,x_-)$ and $(x_+,+\infty)$, where $x_\pm$
($x_-<x_+$) stands for the classical turning points which satisfy
$V_{\rm{eff}}(x_\pm)=\omega^2$. Then, in the region where
$k^{-1}_{\omega l }~d(\ln k_{\omega l })/dx \ll 1$ is satisfied, we
write down
\begin{equation}
     \psi_{\omega l}^{\leftarrow}(x)
        \approx \frac{A^{\leftarrow}_{\omega}}{\sqrt{k_{\omega l}}}
       \left\{
          \begin{array}{lc}
            {\cal T}_{\omega l}^{\leftarrow}  e^{-i(\sigma_{\omega l}-\pi/4)}
            \; (x< 0, |x| \gg 1) \\
            \\
            e^{-i(\rho_{\omega l} +\pi/4)}
            \\
            + {\cal R}_{\omega l}^{\leftarrow}
             e^{i(\rho_{\omega l}+\pi/4)}
            \; (x \gg 1)
          \end{array} \right.
   \label{hhin}
\end{equation}
and
\begin{equation}
     \psi_{\omega l}^{\rightarrow}(x)
       \approx \frac{A^{\rightarrow}_{\omega}}{\sqrt{k_{\omega l}}}
        \left\{
          \begin{array}{l}
             {\cal R}_{\omega l}^{\rightarrow}e^{-i(\sigma_{\omega l}-\pi/4)}
           \\
         +\; e^{i(\sigma_{\omega l}-\pi/4)} \; (x < 0\;, |x| \gg 1) \\
         \\
         {\cal T}_{\omega l}^{\rightarrow}  e^{i(\rho_{\omega l}+\pi/4)}
         \; (x \gg 1)
          \end{array}
        \right.
   \label{hhout}
\end{equation}
where we have defined
$$
\sigma_{\omega l}(x)\equiv
\int^{x}_{x_-}k_{\omega l}(x^\prime)dx'
$$
and
$$
\rho_{\omega l}(x) \equiv
\int^{x}_{x_+}k_{\omega l}(x^\prime)dx'
$$
with
$
k_{\omega l}(x) \equiv 2M \sqrt{\omega^2 -V_{\rm{eff}}(x)}
$.
Here the normalization constants $A^{\alpha}_{\omega}$ are
determined by an asymptotic fitting between
Eqs.~(\ref{68})-(\ref{51}) and Eqs.~(\ref{hhin})-(\ref{hhout}),
respectively. As a result, we obtain
$|A^{\leftarrow}_{\omega}|=|A^{\rightarrow}_{\omega}|=
\sqrt{M/2\omega}$.

Now, we must analyze the case $V_{\rm{eff}}(x) > \omega^2$, which
occurs in the interval $(x_-,x_+)$. In this region,
$\psi^\alpha_{\omega l}(r)$ can be cast in the form
\begin{equation}
\psi_{\omega l}^{\leftarrow}(x)
\approx -i
\frac{A^{\leftarrow}_{\omega}}{ \sqrt{\kappa_{\omega l}}}
e^{-\xi_{\omega l}}
\label{hin}
\end{equation}
and
\begin{equation}
\psi_{\omega l}^{\rightarrow}(x)
\approx -i
\frac{A^{\rightarrow}_{\omega}}{ \sqrt{\kappa_{\omega l}}}
e^{(\Theta_{\omega l}+\xi_{\omega l})}
\label{hout}
\end{equation}
assuming $\kappa^{-1}_{\omega l }~d(\ln \kappa_{\omega l })/dx \ll 1$,
where we have defined
$$
\xi_{\omega l}(x)\equiv \int^{x_+}_{x} \kappa_{\omega l}(x^\prime)
dx^\prime
$$
and
$$
\Theta_{\omega l} \equiv - \int^{x_+}_{x_-} \kappa_{\omega l}(x) dx
$$
with
$\kappa_{\omega l}(x) \equiv 2M \sqrt{V_{\rm{eff}}(x)-\omega^2}$.
The quantity $\Theta_{\omega l}$ is the well-known barrier factor
and is associated with the transmission coefficient by
\begin{equation}
|{\cal T}_{\omega l}^{\alpha}|^2 \approx e^{2 \Theta_{\omega
l}}. \label{barrier}
\end{equation}
As a consequence, $|{\cal R}_{\omega l}^{\alpha}|^2 \approx 1-e^{2
\Theta_{\omega l}}$.

\subsection{Numerical calculation}
\label{sec:numerical}

Now, in order to plot particle and energy emission rates in the
whole frequency range, a numerical calculation procedure is in
order. Briefly speaking the numerical method consists of
solving Eq.~(\ref{53}) for the left- and right-moving
radial functions $ \psi^\leftarrow_{\omega_0 l}(r)$
and $\psi^\rightarrow_{\omega_0 l}(r)$ with asymptotic boundary
conditions compatible with Eqs.~(\ref{68}) and~(\ref{51}),
respectively. We address to Ref.~\cite{CHM2CQG} for more detail.

\section{Results}
\label{sec:results}
\begin{figure}[t]
\includegraphics[width=8cm]{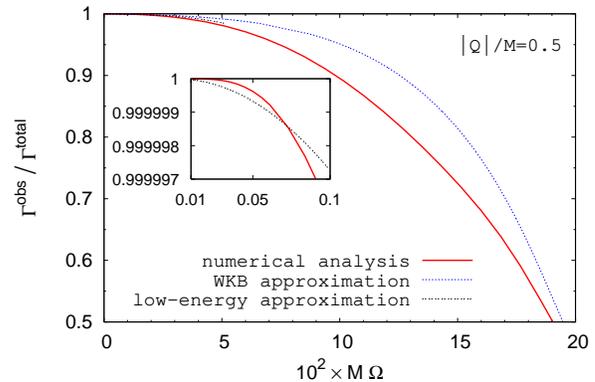}
\caption{We plot the rate of particles which reach
asymptotic observers as a function
of $\Omega$ for $|Q|/M=0.5$. The sum in
Eq.~(\ref{Gammaobs}) is taken up to $l=5$. It is seen that
the WKB and the low-energy approximation give good results
in the proper regions, as expected. It is interesting to note
that for $R_S \approx r_{\rm ph}$ ($\Omega M \approx 0.2$) 
about half of the emitted particles are lost inside the hole.
}
\label{rfracinfr}
\end{figure}
\begin{figure}[b]
\includegraphics[width=8cm]{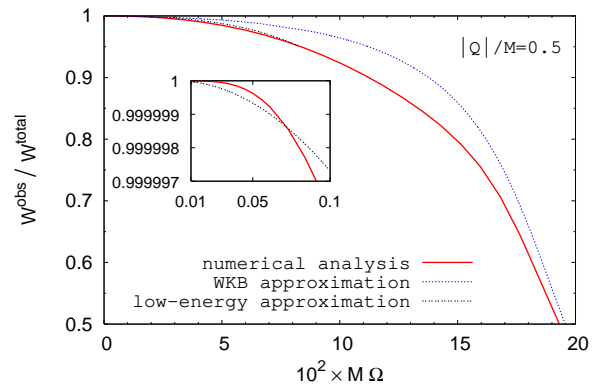}
\caption{The power observed at infinity is plotted as a function
of $\Omega$ for $|Q|/M=0.5$.
}
\label{pfracinfr}
\end{figure}
In Figs.~\ref{crin}  and~\ref{crout} we show the particle emission
rates for $l=m=5$, and  $\alpha = \leftarrow$ and~$\rightarrow$,
respectively. We also display a zoom for  $M \Omega \ll 1$. We can
see the good approximation provided by our low-energy formulas,
which are applicable when the source is in circular orbits far away
from the horizon. In the same token, the results obtained using the
WKB approximation reproduce very well the curves for $R_S \approx
r_{\rm ph}$, i.e., when the source is close to the innermost
timelike geodesic circular orbit. This is the region where most
emitted particles are high-energy ones. This is also convenient to
notice that the WKB approximation reproduces  most qualitative
aspects of the exact numerical calculation. The WKB approximation is
specially good for large angular momentum quantum numbers $l$, as
expected. This is very handy when dealing with large $l$ solutions
since the WKB method requires comparatively modest computational
resources in contrast to the full numerical procedure. In
Figs.~\ref{lowresponse} and~\ref{highresponse} we analyze in more
detail the low- and high-energy particle emission regions by using
the proper formulas, namely, Eq.~(\ref{Gammaem}) with
Eqs.~(\ref{uI})-(\ref{uII}) and Eqs.~(\ref{hin})-(\ref{hout}),
respectively. We note that $\Gamma_{\leftarrow \omega_0 l}$ is
typically larger than $\Gamma_{\rightarrow \omega_0 l}$ except for
$R_S \approx r_{\rm ph}$ and that the larger the $l$  the smaller
the $\Gamma_{\alpha \omega_0 l}$. (For a fixed $l$ the larger the
$m$ the larger the contribution provided that $l+m$ is even.)
Moreover the presence of charge in the black hole tends to damp
$\Gamma_{\alpha \omega_0 l}$. In Figs.~\ref{lowpower}
and~\ref{highpower} we analyze in more detail the radiated power in
the regions where the source is far away from the horizon and close
to the innermost timelike geodesic orbit by using Eq.~(\ref{Wem})
with Eqs.~(\ref{uI})-(\ref{uII}) and Eqs.~(\ref{hin})-(\ref{hout}),
respectively. Far away from the hole the leading contribution to the
power comes from the mode with $l=m=1$, while for $R_S \approx
r_{\rm ph}$ this will depend on the source angular velocity
$\Omega$. In Figs.~\ref{rfracinfr} and~\ref{pfracinfr} we plot the
particle emission rate and corresponding power which reach
asymptotic observers for $|Q|/M=0.5$, respectively. It is worth
noting that for $R_S \approx r_{\rm ph}$ about half of the emitted
particles are absorbed by the hole. The WKB and the low-energy
approximations are in nice agreement with the numerical results.

\section{Final Remarks}
\label{sec:finalremarks}

We have considered the scalar radiation emitted from a monopole
in circular geodesic orbits around a charged black hole.
Emission rates and radiated powers were calculated using exact numerical
and approximate analytical calculations, and shown to be in excellent
agreement with each other in the proper regions.  The net radiation
which reaches asymptotic observers was also investigated and shown
to decrease up to 50\% when the source is close to the innermost
geodesic circular orbit. Because we have only assumed here
particles with small angular momentum and energy in comparison to the
black hole mass and charge, no significant backreaction effects are
expected and, thus, the background spacetime was regarded as fixed.
The possibility of challenging the CCC by the tunneling of particles
with high enough angular momenta through the black hole scattering potential
has been recently discussed~\cite{MS07,H08,RS08}. Here we have offered a 
production mechanism for these particles. Fermionic ones considered in
Ref.~\cite{RS08} can be produced by a similar mechanism.

\acknowledgments

L.\ C. and G.\ M.\ are grateful to Conselho Nacional de Desenvolvimento
Cient\'\i fico e Tecnol\'ogico (CNPq) for partial financial support.
A.\ S.\ and G.\ M.\ would like also to acknowledge partial and full
financial support from Funda\c{c}\~ao de Amparo \`a
Pesquisa do Estado de S\~ao Paulo (FAPESP), respectively.

\end{document}